\documentclass{llncs}
\usepackage[bottom]{footmisc}

\usepackage{makeidx}     
\usepackage{graphicx}    
\usepackage{multicol}    
\usepackage{amssymb}
\usepackage{amsmath}
\usepackage{epstopdf}
\usepackage{color}
\usepackage{subfigure}
\usepackage{hyperref}

\makeindex             
\newenvironment{Sketch}{\par\noindent\textit{Sketch of Proof}\quad}{\hfill\qed\par\smallskip}

\newtheorem{definitn}{Definition}
\newtheorem{thm}{Theorem}
\newtheorem{clm}{Claim}

\begin{document}
\title{Route Distribution Incentives}
\date{}
\author{Joud Khoury\inst{1}, Chaouki T. Abdallah\inst{1}, Kate Krause\inst{2}, and Jorge Crichigno\inst{1}}
\institute{ECE Department, MSC01 1100,\\
1 University of New Mexico, Albuquerque NM 87131\\
\email{\{jkhoury, chaouki, jcrichigno\}@ece.unm.edu}
\and
Economics Department, University of New Mexico\\
1915 Roma NE/Economics Bldg., Albuquerque NM 87131\\
\email{kkrause@unm.edu}
}
\maketitle
\begin{abstract}
We present an incentive model for route distribution in the context of path vector routing protocols and we focus on the Border Gateway Protocol (BGP). BGP is the \textit{de-facto} protocol for interdomain routing on the Internet. We model BGP route distribution and computation using a game in which a BGP speaker advertises its prefix to its direct neighbors promising them a reward for further distributing the route deeper into the network, the neighbors do the same thing with their neighbors, and so on. The result of this cascaded route distribution is an advertised prefix and hence reachability of the BGP speaker.   We first study the convergence of BGP protocol dynamics to a unique outcome tree in the defined game. We then proceed to study the existence of equilibria in the \textit{full information} game considering competition dynamics. We focus our work on the simplest two classes of graphs: 1) the line (and the tree) graphs which involve no competition, and 2) the ring graph which involves competition. 
\end{abstract}
\section{Introduction}
The Border Gateway Protocol (BGP) is a policy-based path vector protocol and is the de-facto protocol for Internet interdomain routing. The protocol's specification~\cite{bgp4} was initially intended to empower domains with control over \textit{route selection}, and \textit{route propagation}. The commercialization of the Internet transformed Autonomous Systems (AS) into economic entities that act selfishly when implementing their internal policies and particularly the decisions that relate to route selection and propagation~\cite{caesar:05}. 
BGP is intrinsically about distributing route information to destinations (which are IP prefixes) to establish paths in the network. Path discovery, or simply discovery hereafter, starting with some destination prefix is the outcome of route distribution and route computation. 

Accounting for and sharing the cost of discovery is an interesting problem whose absence from current path discovery schemes has led to critical economic and scalability concerns. As an example, the BGP control plane functionality is oblivious to cost. A node (BGP speaker) that advertises a provider-independent prefix (identifier) does not pay for the cost of being discoverable.  Such a cost, which may be large given that the prefix is maintained at every node in the Default Free Zone (DFZ), is paid by the rest of the network. For example, Herrin~\cite{herrin_web} has preliminarily analyzed the non-trivial cost of maintaining a BGP route. Such incentive mismatch in the current BGP workings is further exacerbated by provider-independent addressing, multi-homing, and traffic engineering practices~\cite{iabreport:07}.  Given the fact that the number of BGP prefixes in the global routing table (or RIB) is constantly increasing at a rate of roughly $100,000$ entries every $2$ years and is expected to reach a total of $388,000$ entries in $2011$~\cite{potaroo:08}, has motivated us to devise a model that accounts for distribution incentives in BGP. 

A large body of work has focused on choosing the right incentives given that ASes are self-interested, utility-maximizing agents. While exploring incentives, most previous work has ignored the control plane incentives~\footnote{In this paper, we use the term``control plan'' to refer only to route prefix advertisements (not route updates) as we assume that the network structure is static.} (route advertisement/distribution) and has instead focused on the forwarding plane incentives (e.g. transit costs). One possible explanation for this situation is based on the following assumption: a node has an incentive to distribute routes to destinations since the node will get paid for transiting traffic to these destinations, and hence route distribution is ignored as it becomes an artifact of the transit process. We argue that this assumption is not economically viable by considering the arrival of a new customer (BGP speaker). While the servicing edge provider makes money from transiting the new customer's traffic to the customer, the middle providers do not necessarily make money while still incurring the cost to maintain and distribute the customer's route information. 
In this work, we separate the control plane incentives (incentives to distribute route information) from the forwarding plane incentives (incentives to forward packets) and use game theory to model a BGP distribution game. The main problem we are interested in is how to allow BGP prefix information to be distributed globally while aligning the incentives of all the participating agents?
\paragraph{A Simple Distribution Model}
We synthesize many of the ideas and results from~\cite{griffin_icnp:99,kleinberg:05,FRS:06,levin:08} into a coherent model for studying BGP route distribution incentives. Influenced by the social network query propagation model of Kleinberg and Raghavan~\cite{kleinberg:05}, we use a completely distributed model in the sense that it does not assume a central bank (in contrast to previous work on truthful mechanisms~\cite{nisan:07}).
A destination $d$ advertises its prefix and wishes to invest some initial amount of money $r_d$ in order to be globally discoverable (or so that the information about $d$ be globally distributed). Since $d$ may distribute its information to its direct neighbors only, $d$ needs to provide incentives to get the information to propagate deeper into the network.  Therefore, $d$ must incentivize its neighbors to be distributors of its route, who then incentivize their neighbors to be distributors, and so on. 
While we take BGP as the motivating application, we are interested in the general setting of distributing a good to a set of agents. Agents are located on a network and trade may only occur between directly connected agents. Prices are chosen strategically and the agents are rewarded by volume of sales. 

\paragraph{Our Results }
A general model for studying BGP was originally defined by Griffin et. al \ in~\cite{griffin_icnp:99} and later by Levin et. al\ in \cite{levin:08}. In section~\ref{game}.   We build upon this general model to define the \textit{BGP distribution game} and the main goal of this paper is to study the existence of equilibria in the defined game. 
Studying the equilibria for arbitrary graph structures is not an easy problem given the complexity of the strategic dependencies and the competition dynamics. As we are not aware of general existence results that apply to our game, we initially focus on the simplest two classes of graphs: 1) the line (and the tree) graphs which involve no competition, and 2) the ring graph which involves competition. We assume \textit{full information} as we are interested in studying the existence question initially rather than how the players would arrive at the equilibrium.\\  
We show that a subgame perfect equilibrium always exists for the game induced on the line graph (and on the tree), while no such equilibrium exists for the game induced on the ring graph due to oscillation of \textit{best-response} dynamics under competition. While the full game does not have a subgame perfect equilibrium, we show that there always exists a Nash equilibrium for a special class of subgames. This requires us to first quantify the \textit{growth of rewards}, or in other words the minimum incentive $r_d$ such that there exists an equilibrium outcome which is a spanning tree (i.e. $d$ is globally discoverable).
\paragraph{Related work }
The Simple Path Vector Protocol (SPVP) formalism~\cite{griffin_icnp:99} develops sufficient conditions for the outcome of a path vector protocol to be stable. 
A respective game-theoretic model was developed by Levin~\cite{levin:08} that captures these conditions in addition to incentives in a game theoretic setting.
Feigenbaum et. al study incentive issues in BGP by considering least cost path (LCP) policies~\cite{FPSS:05} and more general poilicies~\cite{FRS:06}. Our model is fundamentally different from~\cite{FPSS:05} (and other works based in mechanism design~\cite{selwyn:05}) in that the prices are strategic, the incentive structure is different, and we do not assume the existence of a central ``designer'' (or bank) that allocates payments to the players but is rather completely distributed as in real markets. 
The bank assumption is limiting in a distributed setting, and an important question posed in~\cite{FRS:06} is whether the bank can be eliminated and replaced by direct payments by the nodes. A desirable property of our model is that payments are bilateral and may only flow between neighbors where a player $i$ should not be able to send a payment to another player $j$ unless the latter is a direct neighbor. This renders the model more robust to manipulation.

Li et. al~\cite{cuihongli:07} study an incentive model for query relaying in peer-to-peer (p2p) networks based on rewards, upon which Kleinberg et. al~\cite{kleinberg:05} build to model a more general class of trees. In~\cite{kleinberg:05}, Kleinberg and Raghavan allude to a  similar version of our distribution game in the context of query incentive networks. They pose the general question of whether an equilibrium exists for general Directed Acyclic Graphs (DAGs) in the query propagation game. Both of these probabilistic models do not account for competition. While we borrow the basic idea, we address a different problem which is that of route distribution versus information seeking. 

Finally, our work relates to price determination in network markets with intermediaries (refer to the work by Blume et al.\ \cite{blume:07} and the references therein). A main differentiator of this class of work from other work on market pricing is its consideration of intermediaries and the emergence of prices as a result of strategic behavior rather than competitive analysis or truthful mechanisms. Our work specifically involves the cascading of traders (or distributors) on complex network structures.

\section{The General Game}\label{game}
Reusing notation from~\cite{FRS:06,levin:08}, we consider a graph $G=(V,E)$ where $V$ is a set of $n$ nodes (alternatively termed players, or agents) each identified by a unique index $i=\{1,\ldots,n\}$, and a destination $d$, and $E$ is the set of edges or links. Without loss of generality (WLOG), we study the BGP discovery/route distribution problem for some fixed destination AS with prefix $d$ (as in~\cite{griffin_icnp:99,FRS:06,levin:08}). The model is extendable to all possible destinations (BGP speakers) by noticing that route distribution and computation are performed independently per prefix. The destination $d$ is referred to as the \textit{advertiser} and the set of players in the network are termed \textit{seekers}. Seekers may be distributors who participate in distributing $d$'s route information to other seeker nodes or consumers who simply consume the route (leaf nodes in the outcome distribution tree).  For each seeker node $j$, Let $P(j)$ be the set of all routes to $d$ that are known to $j$ through advertisements, $P(j) \subseteq \mathcal{P}(j)$, the latter being the set of all simple routes from $j$. The empty route $\phi \in \mathcal{P}(j)$. Denote by $R_j \in P(j)$ a simple route from $j$ to the destination $d$ with $R_j=\phi$ when no route exists at $j$, and let $(k,j)R_j$ be the route formed by concatenating link $(k,j)$ with $R_j$, where $(k,j) \in E$. Denote by $B(i)$ the set of direct neighbors of node $i$ and let $next(R_i)$ be the next hop node on the route $R_i$ from $i$ to $d$. Define node $j$ to be an \textit{upstream} node relative to node $i$ when $j \in R_i$. The opposite holds for a \textit{downstream} node. Finally, we use  $r_{next(R_i)}$ to refer to the reward that the upstream parent from $i$ on $R_i$ offers to $i$. 

The general distribution game is as follows: destination $d$  first
exports its prefix (identifier) information to its neighbors promising them a reward $r_d \in \mathbb{Z}^{+}$ 
which directly depends on $d$'s utility of being discoverable. A node $j$ (a player) in turn receives offers from its neighbors where each neighbor $i$'s offer takes the form of a reward $r_{ij}$. A reward $r_{ij}$ that a node $i$ offers to some direct neighbor $j \in B(i)$ is a contract stating that $i$ will pay $j$ an amount that is a function of $r_{ij}$ and of the set of downstream nodes $k$ that decide to route to $d$ through $j$ (i.e. $j \in R_k$ and $R_j=(j,i)R_i$). 
After receiving the offers, player $j$ strategizes by selecting a route among the possibly multiple advertised routes to $d$, say $(j,i)R_i$, and deciding on a reward $r_{jl} < r_{ij}$ to send to each \textit{candidate} neighbor $l \in B(j)$ that it has not received a competing offer from.  Note then that  $r_{lj} < r_{jl}$ where $r_{lj}=0$ means that $j$ did not receive an offer from neighbor $l$.  Node $j$ then pockets the difference $r_{ij}-r_{jl}$. The process repeats up to some depth that is directly dependent on the initial investment $r_d$ as well as on the strategies of the players. We intentionally keep this reward model abstract at this point, but will revisit it later in the discussion when we define more specific utility functions. 
Clearly in this model, we assume that a player can strategize per neighbor, presenting different rewards to different neighbors. This assumption is based on the autonomous nature of the nodes and the current practice in BGP where policies may differ significantly across neighbors (as with the widely accepted Gao-Rexford policies~\cite{AS_gao} for example).

\paragraph{Assumptions}
To keep our model tractable, we take several simplifying assumptions. In particular, we assume that:
\begin{enumerate}
\item the graph is at steady state for the duration of the game i.e. we do not consider topology dynamics; 
\item the advertiser $d$ does not differentiate among the different players (ASes) in the network i.e. the ASes are indistinguishable to $d$. 
\item the advertised rewards are integers and are strictly decreasing with depth i.e. $r_{ij} \in \mathbb{Z}^{+}$ and $r_{ij} < r_{next(R_i)}, \forall \ i,j$. We let $1$ unit be the cost of distribution (a similar assumption was taken in~\cite{kleinberg:05} to avoid the degenerate case of never running out of rewards, referred to as ``Zeno's Paradox'');
\item a node that does not participate will have a utility of zero; 
\item finally, our choice of the utility function isolates a class of policies which we refer to as the Highest Reward Path (HRP). As the name suggests, HRP policies incentivize players to choose the path that promises the highest reward. Such class of policies may be defined more generally to account for more complex cost structures as part of the decision space~\footnote{Metric based policies could be modeled with HRP by fixing one of the players' decisions. For example, fixing $r_{ij}=r_{next(R_i)}-1$, $\forall i, j$ results in hop count metric; or alternatively setting $r_{ij}=r_{next(R_i)}-c_i$, where $c_i$ is some local cost to the node results in Least Cost Path (LCP) policy~\cite{FRS:06}, etc.}. 
We assume for the scope of this work that transit costs are extraneous to the model. 
This is a restrictive assumption given that BGP allows for arbitrary and complex policies  that are generally modeled with a valuation or preference function over the different routes to $d$ (check~\cite{griffin_icnp:99,FRS:06}). 
\end{enumerate}
\paragraph{Strategy Space:} Given a set of advertised routes $P(i)$ where each route $R_i \in P(i)$ is associated with a promised reward $r_{next(R_i)} \in \mathbb{Z}^+$, a \textit{pure strategy} $s_i \in S_i$ of an autonomous node $i$ comprises two decisions: 
\begin{itemize}
\item After receiving offers from neighboring nodes, pick a single ``best'' route $R_i \in P(i)$ (where ``best'' is defined shortly in Theorem~\ref{clm:1}); 
\item Pick a reward vector $r_i = [r_{ij}]_j$ promising a reward $r_{ij}$ to each candidate neighbor $j$ (and export route and reward to respective candidate neighbors). 
\end{itemize}
A strategy profile $\mathbf{s}=(s_1,\ldots,s_n)$ and a reward $r_d$ define an outcome of the game~\footnote{We abuse notation hereafter and we refer to the outcome with simply the strategy profile $\mathbf{s}$ where it should be clear from context that an outcome is defined by the tuple $<\mathbf{s}, r_d>$. Notice that a strategy profile may be associated with an outcome if we model $r_d$ as an action. We refrain from doing so to make it explicit that $r_d$ is not strategic.}. Every outcome determines a set of paths to destination $d$ given by $O_d=(R_1,\ldots, R_n)$. A utility function $u_i(\mathbf{s})$ for player $i$ associates every outcome with a real value in $\mathbb{R}$. 
We use the notation $s_{-i}$ to refer to the strategy profile of all players excluding $i$. The Nash equilibrium is defined as follows:
\begin{definitn}\label{def:NE}
A Nash Equilibrium (NE) is a strategy profile $\mathbf{s^*}=(s_1^*,\ldots,s_n^*)$ such that no player can move profitably by changing her strategy, i.e. for each player $i$,
$u_i(s_i^*,s_{-i}^*) \geq u_i(s_i , s_{-i}^*)$, $\forall s_i \in S_i$.
\end{definitn}

\paragraph{Cost:} 
The cost of participation is local to the node and includes for example the cost associated with the effort that a node spends in maintaining the route information~\footnote{A preliminary estimate of this cost is shown by Herrin~\cite{herrin_web} to be \$0.04 per route/router/year for a total cost of at least \$6,200 per year for each advertised route assuming there are around 150,000 DFZ routers that need to be updated.}. 
Other cost factors that depend on the volume of traffic (proportional to the number of downstream nodes in the outcome $O_d$) are more relevant to the forwarding plane and as mentioned earlier in the assumptions, we ignore this cost in the current model. Hence, we simply assume that every player $i$ incurs a cost $c_i$ which is the cost of participating. We assume for the scope of this paper that the local cost is constant with $c_i=c=1$.

\paragraph{Utility:} 
We experiment with a simple class of utility functions which rewards a node linearly based on the number of sales that the node makes. This model incentivizes distribution and potentially requires a large initial investment from $d$. More clearly, define  $N_i(\mathbf{s})=\{j \in V\backslash \{i\} | i \in R_j\}$ to be the set of nodes that pick their best route to $d$ going through $i$ (nodes downstream of $i$) and let $\delta_i(\mathbf{s}) = |N_i(\mathbf{s})|$. 
Let the utility of a node $i$ from an outcome or strategy profile $\mathbf{s}$ be:
\begin{equation}\label{eq:utility}
u_i(\mathbf{s}) = (r_{next(R_i)} - c_i) + \sum_{\{ j | i=next(R_j)\}}{(r_{next(R_i)} - r_{ij})(\delta_j(\mathbf{s}) +1)}
\end{equation}
The first term $(r_{next(R_i)} - c_i)$ of ( \ref{eq:utility}) is incurred by every participating node and is the one unit of reward from the upstream parent on the chosen best path minus the local cost. Based on the fixed cost assumption, we often drop this first term when comparing player payoffs from different strategies since the term is always positive when $c=1$. The second term of ( \ref{eq:utility}) (the summation) is incurred only by distributors and is the total profit made by $i$ where $(r_{next(R_i)} - r_{ij})(\delta_j(\mathbf{s}) +1)$ is $i$'s profit from the sale to neighbor $j$ (which depends on $\delta_j$). 
A rational selfish node will always try to maximize its utility by picking $s_i=(R_i, [r_{ij}]_j)$. 
There is an inherent tradeoff between $(r_{next(R_i)} - r_{ij})$ and $(\delta_j(\mathbf{s}))$ s.t. $i=next(R_j)$ when trying to maximize the utility in Equation (\ref{eq:utility}) in the face of competition as shall become clear later. A higher promised reward $r_{ij}$ allows the node to compete (and possibly increase $\delta_j$) but will cut the profit margin. 
Finally, we  implicitly assume that the destination node $d$ gets a constant marginal utility of $r_d$ for each distinct player that maintains a route to $d$ - the marginal utility of being discoverable by any seeker - and declares $r_d$ truthfully to its direct neighbors (i.e. $r_d$ is not strategic).

\paragraph{ Convergence under HRP:} Before proceeding with the game model, we first prove the following theorem which results in the Highest Reward Path (HRP) policy.
\begin{thm}\label{clm:1}
In order to maximize its utility, node $i$ must always pick the route $R_i$ with the highest promised reward i.e. such that $r_{next(R_i)} \geq r_{next(R_l)}, \forall \ R_l \in P(i)$.
\end{thm}
The proof of Theorem~\ref{clm:1} is given in Appendix~\ref{appendix:clm1}. The theorem implies that a player could perform her two actions sequentially, by first choosing the highest reward route $R_i$, then deciding on the reward vector $r_{ij}$ to export to its neighbors. Thus, we shall represent player $i$'s strategy hereafter simply with the rewards vector [$r_{ij}$] and it should be clear that player $i$ will always pick the ``best'' route to be the route with the highest promised reward.  When the rewards are equal however, we assume that a node breaks ties consistently.\\
The question we attempt to answer here is whether the BGP protocol dynamics converge to a unique outcome tree $T_d$ under some strategy profile $\mathbf{s}$.
A standard model for studying the convergence of BGP protocol dynamics was introduced by Griffin et al.\ ~\cite{griffin_icnp:99}, and assumes BGP is an infinite round game in which a \textit{scheduler} entity decides on the \textit{schedule} i.e. which players participate at each round (models the asynchronous operation of BGP). 
The authors devised the ``no dispute wheels'' condition~\cite{griffin_icnp:99}, which is the most general condition known to guarantee convergence of possibly ``conflicting'' BGP policies to a unique stable solution (tree). From Theorem~\ref{clm:1}, it may be easily shown that ``no dispute wheels'' exist under HRP policy i.e. when the nodes choose highest reward path breaking ties consistently. This holds since any dispute wheel violates the assumption of strictly decreasing rewards on the reward structure induced by the wheel. Hence, the BGP outcome converges to a unique tree $T_d$~\cite{griffin_icnp:99} under any strategy profile $\mathbf{s}$.  
This result allows us to focus on the existence of equilibria as it directly means that the BGP protocol dynamics converges to a tree under any equilibrium strategy profile.
\subsection{The Static Multi-Stage Game with fixed schedule}
Again, for the scope of this paper, we restrict the analysis of equilibria to the simple line and ring graphs. In order to apply the correct solution concept, we fix the \textit{schedule} of play (i.e. who plays when?) as we formalize shortly. We examine a static version of the full-information game in which each player plays once at a particular stage as determined by its proximity to $d$.
The schedule is based on the inherent order of play in the model: recall that the advertiser $d$ starts by advertising itself and promising a reward $r_d$; the game starts at stage 1 where the direct neighbors of $d$, i.e. the nodes at distance $1$ from $d$, observe $r_d$ and play simultaneously by picking their rewards while the rest of the nodes ``do-nothing''.   At stage $2$, nodes at distance $2$ from $d$ observe the stage 1 strategies and then play simultaneously 
and so on. Stages in this \textit{multi-stage game with observed actions}~\cite{fudenberg-tirole:91} have no temporal semantics. Rather, they identify the network positions which have strategic significance. 
The closer a node is to the advertiser, the more power such a node has due to the strictly decreasing rewards assumption. The key concept here is that it is the \textit{information sets}~\cite{fudenberg-tirole:91} that matter rather than the time of play i.e. since all the nodes at distance  $1$ from $d$ observe $r_d$ before playing, all these nodes belong to the same information set whether they play at the same time or at different time instants. We refer to a single play of the multi-stage game as the \textit{static} game. 
We resort to the multi-stage model (the fixed schedule) on our simple graphs to eliminate the synchronization problems inherent in the BGP protocol and to focus instead on the existence of equilibria. 
By restricting the analysis to the fixed schedule, we do not miss any equilibria. This is due to the fact that the fixed schedule is only meant to replace the notion of ``fair and infinite schedule''~\cite{griffin_icnp:99} with a more concrete order of play. The resulting game always converges in a single play for any strategy profile, and the outcome tree is necessarily one of shortest-paths (in terms of number of hops). 
The main limitation of this model however is that it can not deal with variable costs $c_i$ for which the outcome (HRP tree) might not be a shortest-path tree.
 
Formally, and using notation from~\cite{fudenberg-tirole:91}, each player $i$ plays only once at stage $k>0$ where $k$ is the distance from $i$ to $d$ in number of hops.   At every other stage, the player plays the ``do nothing'' action. The set of player actions at stage $k$ is the stage-$k$ action profile, denoted by $a^k=(a_1^k,,\ldots, a_n^k)$. Further, denote by $h^{k+1}=(r_d, a^1, \ldots, a^k)$, the \textit{history} at the end of stage $k$ which is simply the initial reward $r_d$ concatenated with the sequence of actions at all previous stages. We let $h^1=(r_d)$. Finally, $h^{k+1} \subset H^{k+1}$ the latter being the set of all possible stage-$k$ histories. When the game has a finite number of stages, say $K+1$, then a terminal history $h^{K+1}$ is equivalent to an outcome of the game (which is a tree $T_d$) and the set of all outcomes is $H^{K+1}$.\\ 
The pure-strategy of player $i$ who plays at stage $k>0$ is a function of the history and is given by $s_i:H^k\rightarrow \mathbb{R}^{m_i}$ where $m_i$ is the number of direct neighbors of player $i$ that are at stage $k+1$ (implicit here is that a player always picks the highest reward route). 
Starting with $r_d$ (which is $h^1$), it is clear how the game produces actions at every later stage based on the player strategies resulting in a terminal action profile or outcome. Hence, given $r_d$, an outcome in $H^{K+1}$ may be associated with every strategy profile $\mathbf{s}$, and so the definition of Nash equilibrium (Definition (\ref{def:NE})) remains unchanged. Finally, it is worthwhile noting that the ``observed actions'' requirement (where a player observes the full history before playing) is not necessary for our results in the static game as we shall see in the construction of the equilibrium strategies. Keeping this requirement in the model allows us to classify the play from some stage onward, contingent on a history being reached as a subgame in its own right as we describe next.
\begin{definitn}\cite{fudenberg-tirole:91}
A \textit{proper subgame} of a full game is a restriction of the full game to a particular history. The subgame inherits the properties of the full game such as payoffs and strategies while simply restricting those to the history.
\end{definitn}
In our game, each stage begins a new subgame which restricts the full game to a particular history. For example, a history $h^k$ begins a subgame $G(h^k)$ such that the histories in the subgame are restricted to $h^{k+1}=(h^k, a^k)$, $h^{k+2}=(h^k, a^k, a^{k+1})$, and so on. 
\begin{definitn}\cite{fudenberg-tirole:91}
A strategy profile $\mathbf{s^*}=(s_1^*,\ldots,s_n^*)$ is a \textit{subgame-perfect equilibrium} if it is a Nash equilibrium for every proper subgame of the full game. 
\end{definitn}
Hereafter, the general notion of equilibrium we use is the Nash equilibrium and we shall make it clear when we generalize to subgame perfect equilibria. We are only interested in pure-strategy equilibria~\cite{fudenberg-tirole:91} and in studying the existence question as the incentive $r_d$ varies. 
We now proceed to study the equilibria on special networks.

\section{Equilibria on the Line Graph, the Tree, and the Ring Graph}\label{sec:equilibria}
In the general game model defined thus far, the tie-breaking preferences of the players is a defining property of the game, and every outcome (including the equilibrium) depends on the initial reward/utility $r_d$ of the advertiser. 
\begin{figure}[htbp]
\centering
\subfigure[]{\hspace{0.6in}
\includegraphics[scale=0.35]{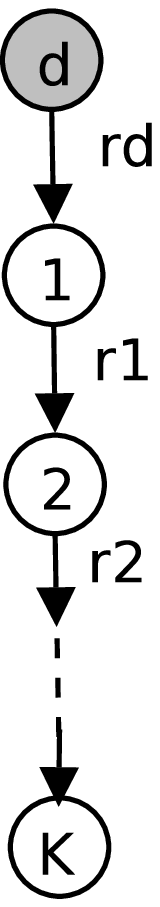}\hspace{0.6in}
\label{fig:line}
}
\subfigure[]{
\includegraphics[scale=0.35]{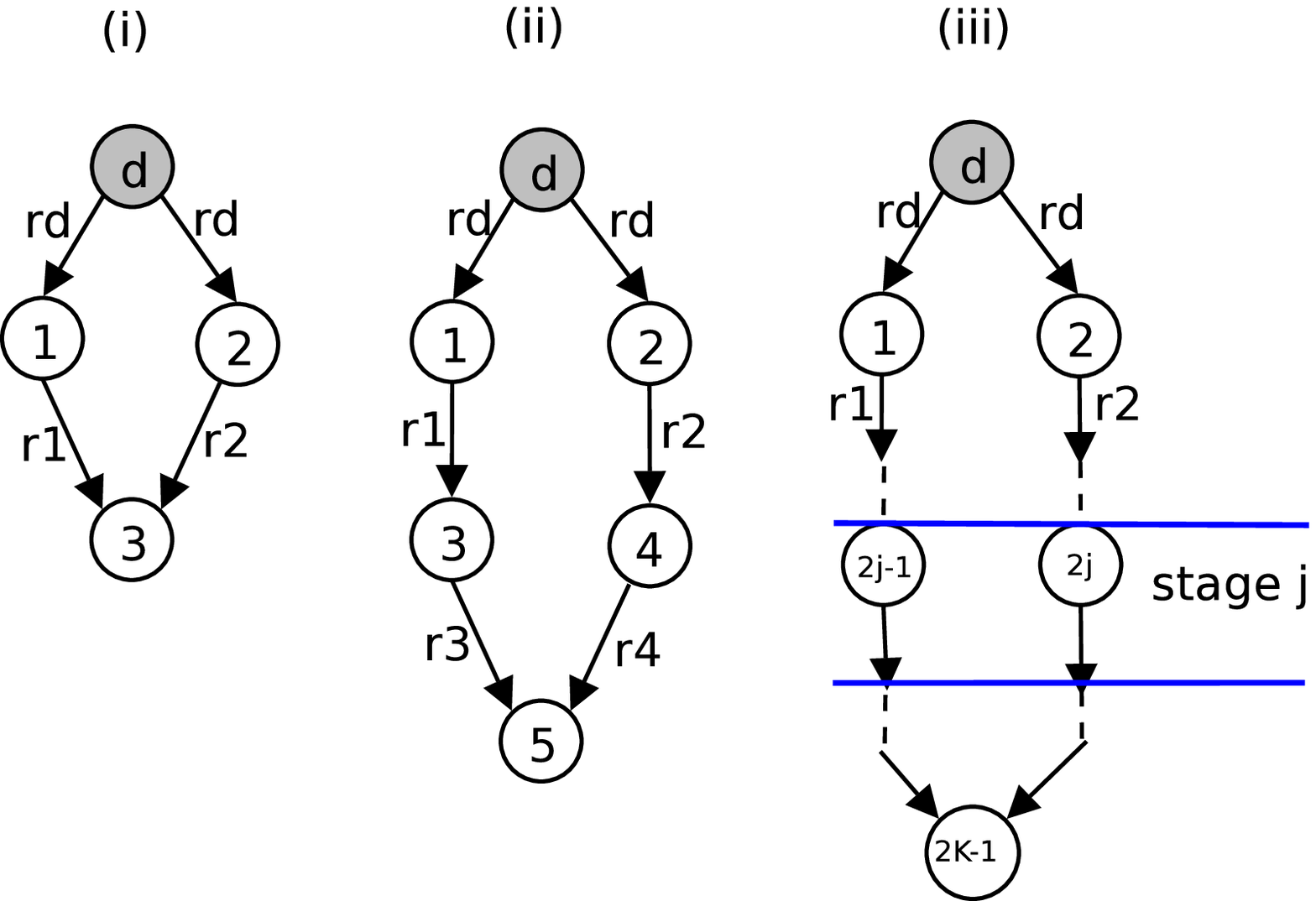}
\label{fig:ring}
}
\label{fig:games}
\caption[]{\subref{fig:line} Line graph: a player's index is the stage at which the player plays; $d$ advertises at stage $0$; $K=n$; \subref{fig:ring} Ring graph with even number of players: (i) $2$-stage game, (ii) $3$-stage game, and general (iii) $K$-stage game.}
\end{figure}
In the same spirit as~\cite{kleinberg:05} we inductively construct the equilibrium for the line graph (simply referred to as the line hereafter) of Figure~\ref{fig:line} given the utility function of Equation (\ref{eq:utility}). We present the result for the line which may be directly extended to trees. Before proceeding with the construction, notice that for the line, $m_i=1$ for all players except the leaf player since each of those players has a single downstream neighbor. In addition, $\delta_i(\mathbf{s})=\delta_j(\mathbf{s}) + 1, \forall i,j$ where $j$ is $i$'s child ($\delta_i=0$ when $i$ is a leaf). We shall refer to both the player and the stage using the same index since our intention should be clear from the context. For example, the child of player $i$ is $i+1$ and its parent is $i-1$ where player $i$ is the player at stage $i$. Additionally, we simply represent the history $h^{k+1}=(r_{k})$ for $k>0$ where $r_k$ is the reward promised by player $k$ (player $k$'s action).  The strategy of player $k$ is therefore $s_k(h^k)=s_k(r_{k-1})$ which is a singleton (instead of a vector) since $m_i=1$ (for completeness, let $r_0=r_d$). This is a \textit{perfect information} game~\cite{fudenberg-tirole:91} since a single player moves at each stage and has complete information about the actions of all players at previous stages. Hence, backward induction may be used to construct the subgame-perfect equilibrium.\\
We construct the equilibrium strategy $s^*$ inductively as follows: first, for all players $i$, let $s^*_i(x)=0$ when $x \leq c$ (where $c$ is assumed to be $1$). Then assume that $s^*_i(x)$ is defined for all $x<r$ and for all $i$. Obviously, with this information, every player $i$ may compute $\delta_i(x, s_{-i}^*)$ for all $x<r$. This is simply due to the fact that $\delta_i$ depends on the downstream players from $i$ who must play an action or reward strictly less than $r$. Finally, for all players $i$ we let $s^*_i(r)=\arg\max_{x} (r - x)\delta_i(x, s_{-i}^*)$ where $x < r$. 
\begin{thm}\label{clm:line}
The strategy profile $s^*$ is a subgame-perfect equilibrium.
\end{thm}
\begin{Sketch} The proof for the line is straightforward and follows from backward induction by constructing the optimal strategies starting with the last player (player $K$) first, then the next-to-last, and so on up to player $1$. The strategies are optimal for every history (by construction) and given the utility function defined in Equation (\ref{eq:utility}), no player can move profitably. Notice that in general when $r_{next(R_i)} \leq c$, propagation of the reward will stop simply because at equilibrium no player will want a negative utility and will prefer to not participate instead (the case with the leaf player).
\end{Sketch}
The proof may be directly extended to the tree since each player in the tree has a single upstream parent as well and backward induction follows in the same way. On the tree, the strategies of the players that play simultaneously at each stage are also independent. 

\subsection{Competition: the ring}\label{sec:ring} 
As opposed to the line, we present next  a negative result for the ring graph (simply referred to as the ring hereafter). In a ring, each player has a degree of $2$ and $m_i=1$ again for all players except the leaf player. We consider rings with an even number of nodes due to the direct competition dynamics. Figure~\ref{fig:ring} shows the $2$-stage, the $3$-stage, and general $K$-stage versions of the game. In the multi-stage game, after observing $r_d$, players $1$ and $2$ play simultaneously at stage $1$ promising rewards $r_1$ and $r_2$ respectively to their downstream children, and so on. We shall refer to the players at stage $j$ using ids $2j-1$ and $2j$ where the stage of a player $i$, denoted as $l(i)$, may be computed from the id as $l(i)=\lceil \frac{i}{2}\rceil$. For the rest of the discussion, we assume WLOG that the player at stage $K$ (with id $2K-1$) breaks ties by picking the route through the left parent $2K-3$. \\
For the $2$-stage game in Figure~\ref{fig:ring}(i), it is easy to show that an equilibrium always exists in which $s_1^*(r_d) = s_2^*(r_d) = (r_d -1)$ when $r_d > 1$ and $0$ otherwise. This means that player $3$ enjoys the benefits of \textit{perfect competition} due to the Bertrand-style competition~\cite{fudenberg-tirole:91} between players $1$ and $2$. 
The equilibrium in this game is independent of player $3$'s preference for breaking ties. 
We now present the following negative result, 
\begin{clm}\label{clm:ring}
The $3$-stage game induced on the ring (of Figure~\ref{fig:ring}(ii)) does not have a subgame-perfect equilibrium. Particularly, there exists a class of subgames for $h^1=rd > 5$ for which there is no Nash equilibrium.
\end{clm}
\begin{Sketch} The proof makes use of a counterexample. 
Using the backward induction argument, notice first that the best strategy of players $3$ and $4$ is to play a \textit{Bertrand-style competition} as follows: after observing $a^1=(r_1,r_2)$, player $3$ plays $r_3=0$ when $r_1=1$, $r_3=\min(r_1-1,r_2 -1)$ when both $r_1 > 1$ and $r_2 > 1$, and $r_3=1$ when $r_1 > 1$ and $r_2 = 1$. Player $4$ plays symmetrically. Knowing that, players $1$ and $2$ will choose their strategies simultaneously and no equilibria exist for $r_d > 5$ due to oscillation of the best-response dynamics. This may be shown by examining the strategic form game, in normal/matrix form, between players $1$ and $2$ (in which the utilities are expressed in terms of $r_d$).
 We briefly show the subgame for $r_d=6$ and we leave the elaborate proof as an exercise for the interested reader. Figure \ref{fig:ring1} shows the payoff matrix of players 1 and 2 for playing actions $r_1 \in \{2,3\}$ (rows) and $r_2 \in \{1,3\}$ (columns), respectively. The payoff shown is taken to be $u_i=(r_d-r_{ij})\delta_i$ ignoring the first term of Equation (\ref{eq:utility}). The actions shown are the only remaining actions after applying iterated strict dominance i.e. all other possible actions for the players are strictly dominated. 
\begin{figure}[htbp]
\begin{center}
\includegraphics[scale=0.4]{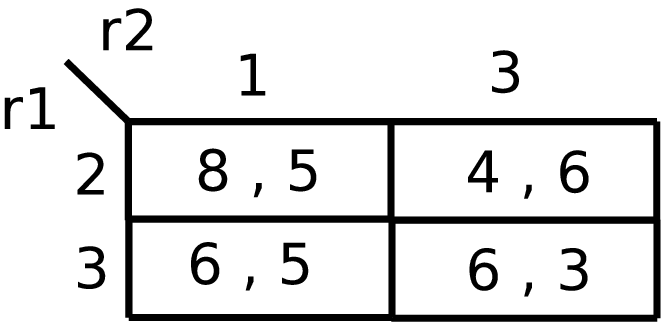}
\caption{The payoff matrix of players $1$ and $2$ for the $3$-stage game on the ring of Figure~\ref{fig:ring}(ii) when $r_d=6$.}
\label{fig:ring1}
\end{center}
\end{figure}
Clearly, no pure strategy Nash equilibria exist. 
The argument could be directly extended to any $r_d > 5$ since player $2$ will still have the incentive to oscillate.
\end{Sketch}
The value $r_d > 5$ signifies the breaking point of equilibrium or the reward at which player $2$, when maximizing her utility $(r_d - r_2)\delta_2$, will always oscillate between competing for $5$ (by playing large $r_2$) or not (by playing small $r_2$). 
Hence, under the linear utility given in Equation (\ref{eq:utility}), an equilibrium does not exist on the simple ring. 
This negative result for the game induced on the $3$-stage ring may be directly extended to the general game for the $K$-stage ring by observing that a class of subgames $G(h^{K-2})$ of the general $K$-stage game are identical to the $3$-stage game. While the full game does not have an equilibrium for $K>2$ stages, we shall show next that there always exists an equilibrium for the special subgame $G(r_d^*)$ (for $h^1=r_d^*$), where the reward $r_d^*$ is the minimum incentive to guarantee that $d$'s route is globally distributed at equilibrium. We define and compute $r_d^*$ next before constructing the equilibrium.

\subsection{Growth of Incentives, and a Special Subgame}\label{sec:specialgame}
We next answer the following question: Find the minimum incentive $r_d^*$, as a function of the depth of the network $K$ (equivalently the number of stages in the multi-stage game), such that there exists an equilibrium outcome for the subgame $G(r_d^*)$ that is a spanning tree. We seek to compute the function $f$ such that $r_d^*=f(K)$. 
First, we present a result for the line, before extending it to the ring. On the line, $K$ is simply the number of players i.e. $K=n$. 
\begin{lemma}\label{prop:fk}
On the line graph, we have $f(0)=0$, $f(1)=1$, $f(2)=2$, and $\forall \ k > 2$
\begin{equation}\label{eq:fk}
f(k)=(k-1)f(k-1)-(k-2)f(k-2)
\end{equation}
\end{lemma}
The proof is presented in Appendix~\ref{app:l1}. Notice that $f(K)$ grows exponentially with the depth $K$ of the line network~\footnote{On the other hand, on complete $d$-ary trees, it may be shown that the function $f(k)=\Theta(k)=\Theta(\log_d{n})$ for $d \geq 2$ since the number of players, and hence $\delta_i$, grows exponentially with depth $K$. 
These growth results on the line graph and the tree seem parallel to the result of Kleinberg and Raghavan~\cite{kleinberg:05} (and the elaboration in~\cite{esteban:07}) which states that the reward required by the root player in order to find an answer to a query with constant probability grows exponentially with the depth of the tree when the branching factor of the tree is $1<b<2$ i.e. when each player has an expected number of offsprings $1<b<2$, while it grows logarithmically for $b>2$.}. 
By subtracting $f(k-1)$ from both sides of the recurrence relation, it may be shown that 
\begin{equation}\label{eq:fkfac}
f(k)-f(k-1)=(k-2)!
\end{equation}
We now revisit the the $K$-stage game of Figure~\ref{fig:ring}(iii) on the ring and we focus on a specific subgame which is the restriction of the full game to $h_1=r_d^*=f(K)$, and we denote this subgame by $G(r_d^*)$. Consider the following strategy profile $\mathbf{s}^*$ for the subgame: players at stage $j$ play $s_{2j-1}^*(h^j)=f(K-j)$, and $s_{2j}^*(h^j)=f(K-j-1), \forall$ $1\leq j \leq K-1$, and let $s_{2K-1}^*(h^K)=0$.
\begin{thm}\label{prop:sg}
The profile $\mathbf{s}^*$ is a Nash equilibrium for the subgame $G(r_d^*)$ on the $K$-stage ring, $\forall \ K > 2$.
\end{thm}
The proof is presented in Appendix~\ref{app:t2}. 
This result may be interpreted as follows: if the advertiser were to play strategically assuming she has a marginal utility of at least $r_d^*$ and is aiming for a spanning tree (global discoverability), then $r_d^*=f(K)$ will be her Nash strategy in the game induced on the $K$-stage ring, $\forall \ K > 2$ (given $\mathbf{s}^*$).

We have shown in Lemma (\ref{prop:fk}) that the the minimum incentive $r_d^*$ on the line (such that there exists an equilibrium spanning tree for the subgame $G(r_d^*)$) as a function of depth $K$ is $r_d^*=f(K)$. We now extend the result to the ring denoting by $f_r(K)$ the growth function for the ring in order to distinguish it from that of the line, $f(K)$.
\begin{corollary} 
On the ring graph, we have $f_r(k) = f(k)$ as given by Lemma (\ref{prop:fk}).
\end{corollary}
\begin{Sketch} We have shown in Theorem (\ref{prop:sg}) that $\mathbf{s}^*$ is a an equilibrium for the subgame $G(r_d^*)$ for $r_d^*=f(K)$ and that the equilibrium is a spanning tree. What remains to show is that $f(K)$ is the minimum incentive required. This follows by isolating the left branch of the ring, which is a line graph that constitutes of player $d$ and all the players with odd identifiers, and using the same argument of Lemma (\ref{prop:fk}) on this branch: an $r_d<f(K)$ allows player $1$ to move profitably by playing an $r_1<f(K-1)$ which violates the spanning tree requirement (by definition of $f$).
\end{Sketch}

\section{Discussion}
The Nash equilibria constructed in this paper are not unique. 
It is additionally well known that in a multi-stage game setting, the Nash equilibrium notion might not be ``credible'' as it could present suboptimal responses to histories that would not occur under the equilibrium profile~\cite{fudenberg-tirole:91}, rendering subgame perfect equilibria more suitable in such circumstances. All the Nash equilibria that we have constructed are credible and are consistent with backward induction for the respective histories of the subgames studied. A distinct aspect of our game is that a player $i$ at stage $k$ may not carry an empty threat to an upstream parent at stage $k-1$, since player $i$'s actions are constrained by the parent's action as dictated by the network structure and the decreasing rewards assumption. In this paper, we have studied the equilibria existence question only. Other important questions include quantifying how hard is it to find the equilibria, and devising mechanisms to get to them. These questions, in addition to extending the results to general network structures and relaxing the fixed cost assumption, are part of our ongoing work.\\
While the distributed incentive model has advantages over centralized mechanisms that rely on a ``designer'', the model might suffer from exponential growth of rewards which could potentially make it infeasible for sparse and large diameter networks. Quantifying the suitability of this model to general network structures and to the Internet connectivity graph specifically requires further investigation. Interestingly, while it is a complex network, the Internet's connectivity graph is a \textit{small-world} network i.e. the average distance between any two nodes on the Internet is small~\cite{barabasi:02}.\\
Finally, we have only considered the setting in which $d$'s marginal utility is constant which seems intuitive in a BGP setting where global reachability is the goal, since every node in the DFZ must keep state information about $d$ or else the latter will be unreachable from some parts of the network. Other economic models that assume the network is a market with elastic demand (based on $d$'s utility) and that determine prices based on demand and supply, are interesting to investigate.  They may even be more intuitive in settings where it makes sense to advertise (or sell) a piece of information to a local neighborhood. 

\bibliographystyle{spmpsci}
\bibliography{../../../bibliography}
\appendix{}
\section{Proof of Theorem~\ref{clm:1}}\label{appendix:clm1}
\begin{proof} The case for $|B(i)|=1$ is trivial. The case for $|B(i)| = 2$ is trivial as well since $i$ will not be able to make a sale to the higher reward neighbor by picking the lower reward offer. Assume that node $i$ has more than $2$ neighbors and that any two neighbors, say $k, l$ advertise routes $R_k, R_l \in P(i)$ s.t. $k = next(R_k), l=next(R_l)$ and $r_{ki} < r_{li}$, and assume that $i$'s utility for choosing route $R_k$ over $R_l$ either increases or remains the same i.e. $u_i^{R_k} \geq u_i^{R_l}$. We will show by contradiction that neither of these two scenarios could happen.\\
\textbf{scenario 1: $u_i^{R_k} > u_i^{R_l}$} From Equation (\ref{eq:utility}), it must be the case that either (case 1) node $i$ was able to make at least one more sale to some neighbor $j$ who would otherwise not buy, or (case 2) some neighbor $j$ who picks $(j,i)R_i$ can strictly increase her $\delta_j(\mathbf{s})$ when $i$ chooses the lower reward path $R_k$. For case 1, and assuming that $r_{ij}$ is the same when $i$ chooses either route, it is simple to show that we arrive at a contradiction in the case when $j \in \{k,l\}$ (mainly due to the strictly decreasing reward assumption i.e. $r_i < r_{next(.)}$); and in the case when $j \notin \{k,l\}$, it must be the case that $j$'s utility increases with $i$'s route choice i.e. $u_j^{(j,i)R_k} > u_j^{(j,i)R_l}$. This contradicts with Equation (\ref{eq:utility}) since w.r.t. $j$, both routes have the same next hop node $i$. The same analogy holds for case 2.\\
\textbf{scenario 2: $u_i^{R_k} = u_i^{R_l}$} Using the same analogy of scenario 1, there must exist at least one neighbor $j$ of $i$ that would buy $i$'s offer only when the latter picks $R_k$, or otherwise node $i$ will be able to strictly increase its utility by picking $R_l$ pocketing more profit. 
\end{proof}

\section{Proof of Lemma~\ref{prop:fk}}\label{app:l1}
\begin{proof}
First, $f(0)=0$, $f(1)=1$ and $f(2)=2$ are trivially true given the utility function of Equation (\ref{eq:utility}). The proof uses induction on the depth of the network. First, for the base case $k=3$, in the $3$-stage line the Nash equilibrium is for player $1$, the player at distance $1$ from $d$, to play $r_1=2$ and for player $2$ to play $r_2=1$ (in every NE, $s_i(1)=0, \ \forall i$). Given $r_d^*=f(3)=3$, the utility of player $1$ is $u_1=(3-2)2 \geq (3-r'_2)\delta'_2, \ \forall r'_2 < 3$. Similarly player $2$ may not move profitably from playing $r_2=1$. \\
Assume $f(x)=(x-1)f(x-1)-(x-2)f(x-2)$ holds $\forall \ x < k$. We construct the $k$-stage game from the $(k-1)$-stage game by adding a node/player between node $d$ and node $1$ in the $(k-1)$-stage game. Notice the player $2$ in the $k$-stage game used to be player $1$ in the $(k-1)$-stage game. By definition of $f$, in the $k$-stage game, when player $1$ plays $r_1=f(k-1)$ then $\delta_1=(k-1)$ and no player $i, 2\leq i \leq k$ may deviate profitably from playing $r_i=f(k-i)$. Here $r_1=f(k-1)$ is the minimum reward to get a $\delta_1=(k-1)$. 
In general, it holds by construction of $f$ that there are $k$ possible outcomes for player $1$, corresponding to the values $\delta_1={0,1,\ldots, k-1}$. For each of these outcomes, we have an action for player $1$, $r_1=f(x)$, which results in the outcome tree corresponding to $\delta_1=x, \forall \ x < k$ and such that no player besides player $1$ may deviate profitably contingent on player $1$ playing $r_1=f(x)$ (In this outcome player $i$ plays $f(x-i+1)$ $\forall \ 2 \leq i \leq n$). 
In order for $\delta_1=k-1$ to be the equilibrium outcome, it must be the case that $r_1=f(k-1)$ maximizes player $1$'s utility given $r_d$ (and hence no player including player $1$ may deviate profitably) i.e. it must be that $\forall \ 2\leq j \leq k$
\begin{equation*}
(r_d - f(k-1))(k-1) \geq (r_d - f(k-j))(k-j)
\end{equation*}
This condition is equivalent to:
\begin{equation}\label{eq:fk1}
(r_d - f(k-1))(k-1) \geq (r_d - f(k-2))(k-2)
\end{equation}
since $(r_d - f(k-2))(k-2) \geq (r_d - f(k-j))(k-j), \forall \ 3 \leq j \leq k$ and for $r_d \geq f(k-1)$. Equation (\ref{eq:fk1}) implies that $r_d \geq (k-1)f(k-1)-(k-2)f(k-2)$. The minimum such incentive is:
\begin{equation}
r_d^* = f(k) = (k-1)f(k-1)-(k-2)f(k-2)
\end{equation}
which is greater than $f(k-1)$ concluding the proof.\qed
\end{proof}

\section{Proof of Theorem~\ref{prop:sg}}\label{app:t2}
\begin{proof}
Notice first that the complete history $h^{K+1}$ which corresponds to $r_d^*$ and $\mathbf{s^*}$ is an outcome that is a spanning tree (each player picks the best route through the upstream parent while the last player $2K-1$ prefers the left parent who is promising a higher reward). We will show that no player $i$ can deviate from playing $s_i^*$ given $s_{-i}^*$ by considering the players at each stage $j, \forall ~2\leq j \leq K-1$ first and then we extend the reasoning to the players at stage $1$. For the players at stage $j$ we show that player $2j-1$ may not deviate profitably from playing $s^*_{2j-1}(h^j)=r_{2j-1}=f(K - j)$ given the strategies of the rest of the players (particularly given $s^*_{2j}(h^j)=r_{2j}=f(K-j -1)$), and the same for player $2j$. Given that $r_{2j} < r_{2j-1}$ (i.e. player $2j$ not competing with player $2j-1$), then by construction of the function $f$, there exists an outcome on the ring such that $\delta_{2j-1}=K-j$ when $r_{2j-1}=f(K-j)$ and $r_{2j} < r_{2j-1}$ (this holds at each stage $ 2 \leq j \leq K-1$ given the tie-breaking preference of player $2K-1$). The utility then to player $2j-1$ of playing $r_{2j-1}=f(K-j)$ is:\\
\begin{eqnarray}
u_{2j-1} &=& (f(K-j+1) - f(K-j))(K-j)\\ 
	&=&(f(K-j+1) - f(K-j-1))(K-j-1) \label{eq:ufk} \\
	&=&(K-j)!
\end{eqnarray}
where the second equality holds by definition of function $f$ (Equation (\ref{eq:fk})) and the third equality holds because $(f(K) - f(K-2))(K-2) = (f(K) - f(K-1) + f(K-1) - f(K-2))(K-2) = ((K-2)! + (K-3)!)(K-2)=(K-1)!$. Given the strategies of the rest of the players, player ${2j-1}$ may not deviate profitably i.e. $u_{2j-1}(f(K-j), s^*_{-(2j-1)}) \geq u_{2j-1}(r', s^*_{-(2j-1)}), \forall ~r' \neq f(K-j)$. This is simply because playing an $r'>f(K-j)$ will strictly decrease $u_{2j-1}$ since $\delta_{2j-1}$ is already maximized ($\delta_{2j-1}=K-j$ in this case), while playing $r'<f(K-j)$ can at best yield player $2j-1$ the same utility when $r'=f(K-j-1)$ (Equation (\ref{eq:ufk})). The same reasoning holds for player $2j$ who may not deviate profitably by playing $r'' \neq f(K-j-1)$. Specifically, any $r'' < f(K-j-1)$ can at best yield player $2j$ the same utility when $r''=f(K-j-2)$, and in order to compete with player $2j-1$ (and possibly increase $\delta_{2j}$) player $2j$ must play $r'' > r_{2j-1} = f(K-j)$ which violates the decreasing rewards assumption. Hence neither player at stage $j$ may deviate profitably for all $ 2 \leq j \leq K-1$. It remains to show that players at stage $1$ may not deviate profitably. First, player $1$ may not deviate profitably using the same argument we used for player $2j-1$ where $j=1$. The utility to player $1$ is $u_1(f(K-1), s^*_{-1}) = (K-1)!$. On the other hand, player $2$ gets the same utility as player $1$ where $u_2(f(K-2), s^*_{-2})=(f(K) - f(K-2))(K-2) = (K-1)!$. In the same way, player $2$ may not deviate profitably since playing any $r'_2 \neq f(K-2)$ may not increase $u_2$ given $s^*_{-2}$. More clearly, in order for player $2$ to compete with player $1$ and possibly increase $\delta_2$ from $K-2$ to $K-1$, player $2$ must play an $r'_2 > f(K-1)$ which in the best case yields a utility $u_2(r'_2, s^*_{-2}) = (f(K) - r'_2)(K-1)< (K-1)!$. Hence, neither player $1$ nor player $2$ may deviate profitably given the strategies of the other players. Finally, the case for player $2K-1$ is trivial. This concludes the proof.\qed
\end{proof}

\end{document}